\documentclass[prl,twocolumn,superscriptaddress,nopacs]{revtex4}

\usepackage{times}
\usepackage{amsmath}
\usepackage{amssymb}
\usepackage{graphicx}
\usepackage{pst-node}

\begin{document}
% You should use BibTeX and apsrev.bst for references
\bibliographystyle{apsrev}
%\preprint{}

\title{Probing Non-Abelian Statistics with Quasiparticle
Interferometry} 
% Optional argument for running titles on pages
%\title[]{}


\author{Parsa Bonderson}
%\email[]{pbonders@theory.caltech.edu}
%\thanks{}
\affiliation{California Institute of Technology, Pasadena, CA 91125}
\author{Kirill Shtengel}
%\email[]{shtengel@physics.ucr.edu}
%\thanks{}
\affiliation{Department of Physics, University of California,
Riverside, CA 92521}
\affiliation{California Institute of Technology, Pasadena, CA 91125}
\author{J. K. Slingerland}
%\email[]{joost@microsoft.com}
%\thanks{}
\affiliation{Microsoft Research, Station Q, Kohn Hall, University of
California, Santa Barbara, CA 93106}

\date{\today}

\begin{abstract}
  We examine interferometric experiments in systems that exhibit non-Abelian
braiding statistics, expressing outcomes in terms of the modular S-matrix. In
particular, this result applies to FQH interferometry, and we give a detailed
treatment of the Read-Rezayi states, providing explicit predictions for the
recently observed $\nu=12/5$ plateau.
\end{abstract}

\pacs{%PACS numbers:
71.10.Pm, %Fermions in reduced dimensions (anyons, composite fermions, Luttinger liquid, etc.
73.43.-f, % Quantum Hall effects
05.30.Pr,	% Fractional statistics systems (anyons, etc.)
73.43.Fj, % Novel experimental methods; measurements
73.43.Jn,  % Tunneling
11.25.Hf.	% Conformal field theory, algebraic structures
}
%\maketitle must follow title, authors, abstract and \pacs
\maketitle

Quantum systems in two spatial dimensions allow for exotic exchange
statistics, characterized by a unitary representation of the braid group
\cite{Leinaas77, Wilczek82b}. This representation may be non-Abelian, acting
on a multi-dimensional internal Hilbert space \cite{Goldin85,Froehlich88}. So
far, experimental evidence for the existence of such anyonic statistics has
only recently been found in the (Abelian) Laughlin states of Fractional
Quantum Hall (FQH) systems \cite{Camino05a,Camino05b}. However, the prospect
of non-Abelian statistics is far more exciting, especially in light
of its potential application in topologically fault-tolerant
quantum computation \cite{Kitaev03,Freedman02a}. There are currently several
observed FQH
states, at filling fractions $\nu =5/2, 7/2, 12/5$
\cite{Willett87,Pan99,Xia04} (and possibly $\nu=3/8, 19/8$),
%\cite{Willett87,Pan99} and $\nu =12/5$ \cite{Xia04} \cite{Lilly99} for 7/2?
that are expected to possess non-Abelian statistics. 
%For recent experimental data and references, see \cite{Eisenstein02}. 
Numerical studies \cite{Morf98,Rezayi00,Read99} suggest that the
$\nu=5/2,12/5$ states should be described respectively by the Moore-Read (MR)
state \cite{Moore91} and the $k=3, M=1$ Read-Rezayi (RR) state \cite{Read99}.
Clearly, as experimental capabilities progress, it becomes increasingly
important to understand how to probe and correctly identify the
braiding statistics of quasiparticles. In this Letter, we explain how, for any
system described by a topological quantum field theory (TQFT) in the infrared
limit (e.g. FQH systems), knowledge of modular S-matrices may be used to
extract this information from interferometry experiments, or, inversely, to
predict the outcomes of such experiments. As an example relevant to current
experimental interests, we obtain explicit results for the RR states.

The topological properties of 2D quantum systems with
an energy gap can be described using TQFTs (or ``modular tensor categories''
in mathematicians' terminology, see e.g.
\cite{Preskill-lectures,Kitaev06a,Turaev94,Kassel95}), often
abstracted from
conformal field theories (CFTs, see \cite{DSM} and references therein). Such
an anyon model is defined by (i) a finite set $\mathcal{C}$ of particle types
or ``anyonic charges,'' (ii) fusion
rules specifying how these particle types may combine or split, and (iii)
braiding rules dictating the behavior under exchange of two particles (all
subject to certain consistency conditions). The ``vacuum'' charge
is given the label $\openone$. The anti-particle or ``charge
conjugate'' of a particle type $a$ is denoted $\overline{a}$, and is the
unique charge that can fuse with $a$ to give $\openone$. Fusion of particle
types generalizes the addition of charges or angular momenta and the
(commutative and associative) fusion rules are specified as 
$a\! \times b=\sum_{c \in \,\mathcal{C}}N_{ab}^{c}\,c$,
where the integer $N_{ab}^{c}$ is the dimension of the Hilbert space of
particles of type $a$ and $b$ restricted to have total anyonic charge $c$.
Fusion and braiding can be represented diagrammatically on oriented, labeled
particle worldlines, and are unaffected by smooth deformations in which the
lines do not intersect. Charge conjugation is represented by reversal of
wordline orientation. We will refer to only one braiding relation, known as
the modular S-matrix, defined by the following diagram:
\begin{equation}
\label{eq:S-matrix}
S_{ab}=\frac{1}{D}
\pspicture[0.5](2.4,1.3)
  \psarc[linewidth=1pt,linecolor=black,arrows=<-,arrowscale=1.5,
arrowinset=0.15] (1.6,0.7){0.5}{165}{363}
  \psarc[linewidth=1pt,linecolor=black] (0.9,0.7){0.5}{0}{180}
  \psarc[linewidth=1pt,linecolor=black,border=3pt,arrows=->,arrowscale=1.5,
arrowinset=0.15] (0.9,0.7){0.5}{180}{375}
  \psarc[linewidth=1pt,linecolor=black,border=3pt] (1.6,0.7){0.5}{0}{160}
  \psarc[linewidth=1pt,linecolor=black] (1.6,0.7){0.5}{155}{170}
  \rput[bl]{0}(0.15,0.3){$a$}
  \rput[bl]{0}(2.15,0.3){$b$}
  \endpspicture
.
\end{equation}
Here $D=\sqrt{\sum\nolimits_{a}d_{a}^{2}}=1/S_{\openone\openone}$ is the
total quantum dimension, where $d_{a}$, the quantum dimension of particle type
$a$, is the value of a single loop of that type,%
\begin{equation}
d_{a}=
  \pspicture[0.5](1.5,1.3)
 
\psarc[linewidth=1pt,linecolor=black,arrows=->,arrowscale=1.5,arrowinset=0.15]
(0.7,0.7){0.5}
	{0}{373}
  \rput[bl]{0}(1.25,0.3){$a$}
  \endpspicture
  =D S_{{\openone}a}\,.
\end{equation}%
Some useful properties of the S-matrix are
\begin{equation}
\label{sprops}
S_{ab}=S_{ba}=\overline{(S^{-1})}_{ab}=\overline{S}_{\overline{a}b}\,.
\end{equation}
The importance of the S-matrix becomes clear when one envisions interferometry
experiments for these systems in which a particle has two possible paths that
it may take around another particle, the two paths combining to form a closed
loop. This is typical of
Mach-Zender, two-slit, FQH two-point-contact, etc.{} experiments
\cite{Verlinde91,Lo93,Bais92,Bais93,Chamon97,Fradkin98,Overbosch01,DasSarma05,
Stern06a,Bonderson06a}. In such
experiments, an interference term arises that can be written as
\begin{equation}
\left\langle \Psi _{ab}\right| U_{1}^{-1}U_{2}\left| \Psi _{ab}\right\rangle
=e^{i\alpha _{ab}}\left\langle \Psi _{ab}\right| \mathbb{M}\left| \Psi
_{ab}\right\rangle =e^{i\alpha _{ab}}M_{ab}
\end{equation}%
where $\left| \Psi_{ab}\right\rangle$ is the initial state of particles $a$
and $b$, and $U_{1},U_{2}$ are the unitary evolution operators for the
particle $a$
traveling around the particle $b$ via the two respective paths. It has been
rewritten in terms of the monodromy operator $\mathbb{M}$ that contains only
the contribution from adiabatically transporting particle $a$ around particle
$b$ (i.e. braiding), and a phase $e^{i\alpha_{ab} }$ that absorbs all other
contributions (i.e. it contains the free particle dynamics and the
Aharonov-Bohm phase from a background magnetic flux). For simplicity, we let
the two particles have definite anyonic charge, however it is a
straightforward generalization to allow superpositions of particle type. If
the theory only has Abelian statistics, then $\left| M_{ab}\right| =1$, but
with non-Abelian statistics, $\left| M_{ab}\right|$ can be less than $1$ and
must be calculated by TQFT methods. The braiding term is diagrammatically
represented by winding the worldline of particle $a$ around that of particle
$b$, taking the standard closure (where the worldline of each particle is
closed back on itself in a manner that introduces no additional braiding) and
dividing by the quantum dimension of the two particle types \footnote{This
result is derived for initial states of two uncorrelated particles. Other
initial states and correlations may require more involved arguments along the
lines set out in \cite{Overbosch01}.}. Thus, the resulting monodromy matrix
element can be written entirely in terms of the S-matrix:%
\begin{equation}
\label{eq:MS}
M_{ab}=\frac{1}{d_{a}d_{b}}
  \pspicture[0.5](2.4,1.3)
  \psarc[linewidth=1pt,linecolor=black,arrows=<-,arrowscale=1.5,
arrowinset=0.15] (1.6,0.7){0.5}{165}{363}
  \psarc[linewidth=1pt,linecolor=black] (0.9,0.7){0.5}{0}{180}
  \psarc[linewidth=1pt,linecolor=black,border=3pt,arrows=->,arrowscale=1.5,
arrowinset=0.15] (0.9,0.7){0.5}{180}{375}
  \psarc[linewidth=1pt,linecolor=black,border=3pt] (1.6,0.7){0.5}{0}{160}
  \psarc[linewidth=1pt,linecolor=black] (1.6,0.7){0.5}{155}{170}
  \rput[bl]{0}(0.15,0.3){$a$}
  \rput[bl]{0}(2.15,0.3){$b$}
  \endpspicture
  =\frac{S_{ab}S_{\openone\openone}}{S_{{\openone}a}S_{{\openone}b}}\,.
\end{equation}%
This result is particularly nice because the S-matrix is
typically more readily computable than the complete set of braiding/fusion
rules, and, in fact, has already been computed for most physically relevant
theories. In particular, this applies to the class of theories described by
CFTs generated as products and cosets of Wess-Zumino-Witten
theories, which includes all proposed non-Abelian FQH states (see 
\cite{Froehlich01,Wen91a,Ardonne99,Ardonne02}). In the case of the product of
two CFTs or TQFTs,
the particle types are denoted by pairs of particle labels, one from each
theory, and the S-matrix of the product theory is the tensor product of the
S-matrices of the parent theories. In the ${G}_{k}/{H}_{l}$
coset theory, the new labels are also formed as pairs of labels from the
parent ${G}_{k}$ and ${H}_{l}$ theories, but now there are
branching rules which restrict the allowed pairings. Also, different pairs of
labels may sometimes turn out to represent the same particle type, a
phenomenon described by ``field identifications''. This means that there will
be only one row and column in the S-matrix for any set of identified labels.
Despite these complications, the coset's S-matrix elements are described
by the simple formula \cite{Gepner89}  
\begin{equation}
\label{cosetsmat}
S^{(G/H)}_{(a,p)(b,q)}=c(G,H,k,l) \,S^{(G)}_{ab}\bar{S}^{(H)}_{pq}
\end{equation}%
where $a,b$ and $p,q$ are respectively labels of the ${G}_{k}$ and
${H}_{l}$ theories, and $c(G,H,k,l)$ is an overall normalization
constant that enforces unitarity of the S-matrix (but is
irrelevant in Eq.~(\ref{eq:MS})).

\begin{figure}[hbt]
\includegraphics[width=2.4in]{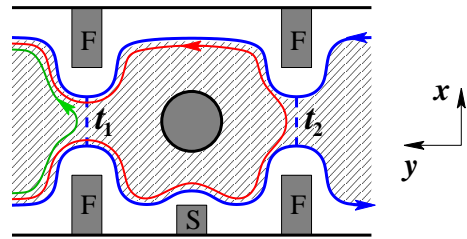}
\caption{A two point-contact interferometer for measuring braiding statistics.
The hatched region contains an incompressible FQH liquid. The front gates
(F) are used to bring the opposite edge currents (indicated by
arrows) close to each other to form two tunneling junctions.  Applying voltage
to the central gate creates an antidot in the middle and controls the number
$n$ of quasiholes contained there. An additional side gate (S) can be used to
change the shape and the length of one of the paths in the interferometer.}
\label{fig:interferometer}
\end{figure}

We now focus specifically on FQH systems, because they are the only physical
systems found to exhibit braid statistics thus far, and represent the most
likely candidates for finding non-Abelian statistics. We consider the
interferometry experiment originally
proposed in \cite{Chamon97} for measuring braiding statistics in the Abelian
FQH states, which was later adopted for the non-Abelian case in
\cite{Fradkin98} and addressed again in the context of the $\nu=5/2$ state in
\cite{Stern06a,Bonderson06a}. A somewhat similar experiment has recently been
implemented to probe the Abelian $\nu=1/3$ state \cite{Camino05a,Camino05b}.
The experimental setup is a two point-contact interferometer composed of a
quantum Hall bar with two front gates on either side of an antidot (see
Fig.~\ref{fig:interferometer}).
Biasing the front gates, one may create constrictions in the
Hall bar, adjusting the tunneling amplitudes $t_{1}$ and $t_{2}$. Tunneling
between the opposite edge currents leads to a 
deviation of $\sigma_{xy}$ from its quantized value, or equivalently,
to the appearance of $\sigma_{xx}$.  By measuring $\sigma_{xx}$ one
effectively measures the interference between the two tunneling paths around
the antidot.  The tunneling amplitudes $t_{1}$ and $t_{2}$ must be kept small,
to ensure that the tunneling current is completely due to quasiholes rather
than higher charge composites \footnote{In the weak tunneling regime, the
tunneling current $I\propto V^{2s-1}$ where $s$ is the scaling
dimension/topological spin of the corresponding fields \cite{Wen92b}. It
follows that the dominant contribution is from the field with lowest scaling
dimension, which in FQH systems is the quasihole.}, and
to allow us to restrict our attention to the lowest order winding term.

In order to be able to influence the resulting interference pattern, we
envision several experimentally variable parameters: (i) the central gate
voltage allowing one to control the number $n$ of quasiholes on the antidot,
(ii) the perpendicular magnetic field $B$, (iii) the back gate voltage
controlling the uniform electron density, and (iv) the side gate that can be
used to modify the shape of the edge inside the interferometric loop
\cite{Stern06a}. The intention is to be able to separately affect the
Abelian Aharonov-Bohm phase and the number of quasiholes on the antidot;
from this point of view having all these controls is redundant, but may be
found beneficial for experimental success.

%%%%%%%%%%%%%%%%%%%%%%%%%%%%%%%%%%%%%%%%%%%%%%%%%%%%%%%%%%%%
%%%%%%%%%%%%%%%%%%%%%%%%%%%%%%%%%%%%%%%%%%%%%%%%%%%%%%%%%%%%

The longitudinal conductivity is proportional to the probability that current
entering the bottom edge will leave through the top edge, which to
lowest order is:%
\begin{eqnarray}
\sigma _{xx} &\propto &\left| t_{1}\right| ^{2}+\left| t_{2}\right| ^{2}+2%
\text{Re}\left\{ t_{1}^{\ast }t_{2}\left\langle \Psi _{ab}\right|
U_{1}^{-1}U_{2}\left| \Psi _{ab}\right\rangle \right\}  \notag \\
&=&\left| t_{1}\right| ^{2}+\left| t_{2}\right| ^{2}+2\left| t_{1}
 t_{2}\right| \left| M_{ab}\right| \cos \left( \beta +\theta
_{ab}\right)\,.
\end{eqnarray}%
In this equation $\beta =\alpha _{ab}+\arg \left( t_{2}/t_{1}\right) $ can be
varied by changing B (keeping the quasihole number fixed), the relative
tunneling phase, and/or the edge shape around the central region. We have
written $M_{ab}=\left| M_{ab}\right|
e^{i\theta _{ab}}$, and will be interested in the elements where the particle
$a$ which tunnels carries the anyonic charge of a quasihole, while the
``particle" $b$ is a composite of $n$ quasiholes on the antidot, carrying an
anyonic charge allowed by the fusion rules \footnote{We expect the states of
the antidot with the same electric charge but different topological spins to
be non-degenerate (with energy difference scaling as $L^{-1}$ for an antidot
of size $L$) and to have different charge distributions, ruling out
the possibility of their superposition over extended time.}. 

We will now apply this formalism to the RR states and,
in particular, look more closely at the $k=3,M=1$ case, which, by a 
particle-hole transformation (which generally inverts the statistics and has
the effect of conjugating the S-matrix), is the expected description of
$\nu=12/5$. The anyon
theory for RR states at $\nu=k/(kM+2)$ can be described
as $\text{RR}_{k,M}=\text{U}(1)
_{k,M}\times \text{Pf}_{k}$ where $\text{U}(1) _{k,M}$ is due to the electric
charge and $\text{Pf}_{k}$ represents the $\mathbb{Z}_{k}$-parafermion theory
\cite{Zamolodchikov85,Gepner87}. The $\text{U}(1) _{k,M}$ part of this theory
is a simple Abelian contribution, essentially labeled by integral multiples of
the charge/flux unit $\left(\frac{e}{kM+2},\frac{2\pi}{ke}\right) $, where
$-e$ is the electron charge and $\frac{2\pi }{e}=\Phi_{0}$ is the magnetic
flux quantum (in units $\hbar=c=1$). The fusion rules for these labels are
just addition of charge/flux and the S-matrix is $S_{n_a n_b} =  e^{i n_{a}
n_{b}
\frac{2\pi }{k\left( kM+2\right) }}/\sqrt{k(kM+2)}$.

The $\text{Pf}_{k}$ part of this theory requires more explanation (for a
discussion of its braiding, see \cite{Slingerland01}). Essentially,
we use that the theory is equivalent to the coset
${\text{SU}(2)}_k/{\text{U}(1)}_k$. As a consequence, the
$\mathbb{Z}_{k}$-parafermion sector's anyonic charge can be labeled by the
corresponding CFT fields
$\Phi_{\lambda }^{\Lambda }$, where $\Lambda \in \left\{ 0,1,\ldots
,k\right\}$
and $\lambda \in \mathbb{Z}$, subject to the identifications $\Phi
_{\lambda }^{\Lambda }=\Phi _{\lambda +2k}^{\Lambda }=\Phi _{\lambda
-k}^{k-\Lambda }$ and the restriction $\Lambda +\lambda \equiv 0\left(
\text{mod}2\right) $ (giving a total of $\frac{1}{2}k\left( k+1\right) $
fields).
The fusion rules for this sector are (as for the CFT fields):%
\begin{equation}
\Phi _{\lambda _{a}}^{\Lambda _{a}}\times \Phi _{\lambda _{b}}^{\Lambda
_{b}}=\sum\limits_{\Lambda =\left| \Lambda _{a}-\Lambda _{b}\right| }^{\min
\left\{ \Lambda _{a}+\Lambda _{b},2k-\Lambda _{a}-\Lambda _{b}\right\} }\Phi
_{\lambda _{a}+\lambda _{b}}^{\Lambda }\,.
\end{equation}%
Their quantum dimensions are%
\begin{equation}
d_{\Phi _{\lambda }^{\Lambda }} =
\sin \left( \frac{\left( \Lambda +1\right) \pi }{k+2}\right) \mbox{\LARGE$/$}
\sin \left( \frac{\pi }{k+2}\right) \,.
\end{equation}%
Special fields in this theory are the vacuum ${\openone} \equiv \Phi
_{0}^{0}$, the $\mathbb{Z}_{k}$-parafermions $\psi
_{l} \equiv \Phi _{2l}^{0}$, the primary fields $\sigma _{l} \equiv \Phi
_{l}^{l}$, and the $\mathbb{Z}_{k}$-neutral excitations $\varepsilon _{j} \equiv
\Phi _{0}^{2j}$, where $l=1,\ldots ,k-1$ and $j=1,\ldots ,\left\lfloor \left(
k-1\right)/2\right\rfloor $. {}From (\ref{cosetsmat}) we find that the S-matrix for 
$\text{Pf}_{k}$ is
\begin{equation}
S_{\Phi _{\lambda _{a}}^{\Lambda _{a}}\Phi _{\lambda _{b}}^{\Lambda _{b}}}=%
\frac{\sin \left( \frac{\left( \Lambda _{a}+1\right) \left( \Lambda
_{b}+1\right) \pi }{k+2}\right) }{D\sin \left( \frac{\pi }{k+2}\right) }%
e^{-i\lambda _{a}\lambda _{b}\pi /k}.
\end{equation}%

The $\text{U}(1)_{k,M}$ and $\text{Pf}_k$ sectors combine so that the anyonic
charges in the $\text{RR}_{k,M}$ theory
are (defining a shorthand) $\hat{\Lambda}_{n}\!\!\!\!\equiv \!\!\!\!
\left(\frac{ne}{kM+2},\frac{n2\pi }{ke},\Phi _{n}^{\Lambda _{n}}\right)$,
where we have $n \!\!\!\!\in\!\!\!\! \mathbb{Z}$ and $\Lambda _{n}
\in \left\{0,1,\ldots ,k\right\}$ such that
$\Lambda_{n} + n \equiv 0\left(\text{mod}2\right)$.
Quasiholes carry
charge \mbox{$\hat{1}_{1} =
\left(\frac{e}{kM+2},\frac{2\pi}{ke},\sigma_{1}\right)$}. 
The S-matrix for RR anyons is obtained by multiplying the
S-matrix elements of the two sectors for anyons $a$ and $b$, and 
renormalizing by some overall constant $c$ (which we will not need explicitly):
\begin{equation}
S_{ab}=c\sin \left( \frac{\left( \Lambda _{n_{a}}+1\right) \left( \Lambda
_{n_{b}}+1\right) \pi }{k+2}\right) e^{-in_{a}n_{b}\frac{M\pi }{kM+2}} .
\end{equation}
Since the tunneling current is dominated by quasiholes \footnote{As previously
noted, the particle with lowest topological spin dominates tunneling. For
RR$_{k,M}$ particles with charge $\hat{\Lambda}_{n}$, the topological spin is
given by $s_{\hat{\Lambda}_{n}}=\frac{n^{2}}{2k\left(
kM+2\right)}+\frac{\Lambda \left( \Lambda +2\right)}{4\left( k+2\right)
}-\frac{\lambda ^{2}}{4k}$, where the field identifications must be used to
write $\Phi _{n}^{\Lambda_{n}}=\Phi _{\lambda }^{\Lambda }$ such that
$-\Lambda \leq \lambda \leq \Lambda $ in order to apply this formula. One may
check that all other particles have greater spin than the
quasihole value, $s_{\hat{1}_{1}}=\frac{kM-M+3}{2\left(
k+2\right) \left( kM+2\right) }$. }, we only need the monodromy matrix
elements%
\begin{equation}
M_{\hat{1}_{1},\hat{\Lambda}_{n}} \!\!
= \frac{\cos \left( \frac{\left( \Lambda _{n}+1\right) \pi }{k+2}\right)}
{\cos\left( \frac{\pi }{k+2}\right)} e^{-in\frac{M\pi}{kM+2}} .
\end{equation}
We note that RR$_{2,1}$ is the MR state, and we can easily check
that this exactly matches the results of \cite{Fradkin98,Bonderson06a}.

We now turn to the $\text{RR}_{3,1}$ theory for $\nu =12/5$. The
$\text{Pf}_{3}$
theory has six fields: $\openone$, $\psi _{1}$, $\psi_{2}$, which have
quantum dimension $1$, and $\sigma _{1}$, $\sigma_{2}$, $\varepsilon$, which
have quantum dimension $\phi=2\cos \frac{\pi}{5}=\frac{1+\sqrt{5}}{2}$ (the
golden ratio).
The total quantum dimension is $D=\sqrt{3\left( \phi +2\right) }$
and the S-matrix is%
\begin{equation*}
S=\frac{1}{D}\left[ 
\begin{array}{cccccc}
1 & 1 & 1 & \phi  & \phi  & \phi  \\ 
1 & e^{i\frac{2\pi }{3}} & e^{-i\frac{2\pi }{3}} & \phi e^{-i\frac{2\pi }{3}}
& \phi e^{i\frac{2\pi }{3}} & \phi  \\ 
1 & e^{-i\frac{2\pi }{3}} & e^{i\frac{2\pi }{3}} & \phi e^{i\frac{2\pi }{3}}
& \phi e^{-i\frac{2\pi }{3}} & \phi  \\ 
\phi  & \phi e^{-i\frac{2\pi }{3}} & \phi e^{i\frac{2\pi }{3}} & e^{-i\frac{
\pi }{3}} & e^{i\frac{\pi }{3}} & -1 \\ 
\phi  & \phi e^{i\frac{2\pi }{3}} & \phi e^{-i\frac{2\pi }{3}} & e^{i\frac{
\pi }{3}} & e^{-i\frac{\pi }{3}} & -1 \\ 
\phi  & \phi  & \phi  & -1 & -1 & -1
\end{array}
\right] 
\end{equation*}%
where the columns and rows are in the order: $\openone$, $\psi _{1}$, $\psi
_{2}$, $\sigma _{1}$, $\sigma _{2}$, $\varepsilon$. Quasiholes in the
RR$_{3,1}$
theory have anyonic charge
$\left(\frac{e}{5},\frac{2\pi}{3e},\sigma_{1}\right) $. It is useful to
consider a Bratteli diagram (which has periodicity 6 in $n$) to keep track
of the allowed $\text{Pf}_3$ charge for a corresponding value of $n$:
\begin{equation*}
\begin{array}{ccccccccc}
& 3 & \multicolumn{1}{|l}{\hspace*{2mm} } &\psi _{2} &  & \openone &  &
\psi_{1} &  \\
& 2 & \multicolumn{1}{|l}{\hspace*{2mm} \varepsilon } &  & \sigma _{2}& 
&\sigma_{1}&  & \varepsilon \\ \Lambda _{n} & 1 &
\multicolumn{1}{|l}{\hspace*{2mm} } & \sigma _{1} &  & %
\varepsilon  &  & \sigma _{2} &  \\
& 0 & \multicolumn{1}{|l}{\hspace*{2mm} \openone} &  & \psi _{1} &  & \psi
_{2} &  & \openone \\[4pt] \cline{2-9} &&\multicolumn{1}{l}{}&&&&&&\\[-3mm]
& n\rightarrow & \hspace*{2mm} 0 & 1 & 2 & 3 & 4 & 5 & 6 \\
%&  & \hspace*{2mm}  &  & n &  &  &  &
\end{array}
\end{equation*}
The longitudinal conductivity in the interferometry experiment will be%
\begin{equation}
\label{eq:sigmaxx}
\sigma _{xx}\propto \left| t_{1}\right| ^{2}\!+\left| t_{2}\right|
^{2}\!+2\left| t_{1} t_{2}\right| \left( -\phi ^{-2}\right)
^{N_{\phi }} \! \cos \! \left( \beta +n\frac{4\pi }{5}\right) 
\end{equation}
where $N_{\phi }=1$ if the $n$ quasihole composite on the
antidot has Pf$_3$
charge with quantum dimension $\phi $ (i.e. $\sigma _{1}$, $\sigma _{2}$, or
$\varepsilon $) and $N_{\phi}=0$ if the composite has quantum dimension $1$
(i.e. $\openone$,
$\psi _{1}$, or $\psi _{2}$). Thus, depending on the total Pf$_3$ charge on
the antidot, 
one of two possible conductivity values will be observed. The Pf$_3$ charge
may then be determined by varying $\beta$ (using the side gate) to measure the
interference fringe amplitude, which
is suppressed by a factor of $\phi^{-2} \approx .38$ when $N_{\phi}=1$. This
behavior indicates
the presence of non-Abelian statistics, and distinguishes this state from
other proposals for the same filling
fraction (e.g. composite fermions). To describe $\nu=12/5$,
we apply a particle-hole transformation to $RR_{3,1}$, replacing $S$ with
$\overline{S}$ (hence, $M$ with $\overline{M}$), which changes the sign in
front of $n$ in Eq.~(\ref{eq:sigmaxx}) \footnote{We thank S.~B.~Chung for
bringing this sign change to our attention.}. In more general scenarios where
composites of quasiholes/quasielectrons may be used instead of the single
tunneling quasiholes, identical behavior (up to the phase) will be exhibited
by quantum dimension $\phi$ composites, while quantum dimension $1$ composites
exhibit a single unsuppressed interference pattern at any $n$.

% where $N_{\phi }=1$ if the $n$ quasihole composite on the antidot has Pf$_3$
% charge with quantum dimension $\phi $ (i.e. $\sigma _{1}$, $\sigma _{2}$, or
% $\varepsilon $) and zero if it has quantum dimension $1$ (i.e. ${\openone}$,
% $\psi _{1}$, or $\psi _{2}$). To describe $\nu=12/5$, we 
% apply a particle-hole transformation to $\text{RR}_{3,1}$, which has the
% effect of changing the sign in front of $n$ in Eq.~(\ref{eq:sigmaxx})
% \footnote{We thank S.~B.~Chung for bringing this to our attention.}.
% Thus, we find two possible values
% of conductivity at each $n$, 
% one with an interference term suppressed by a factor of $\phi^{-2} 
% \approx .38$ with respect to the other.

We conclude with a few remarks explaining that despite the relatively
simple nature of these interferometry experiments, they provide a
surprisingly large amount of information about the system being probed. This
is because the experiments essentially measure the S-matrix of the TQFT that
describes the system.  The S-matrix fully determines the fusion rules through
the Verlinde formula \cite{Verlinde88}:
$N_{ab}^{c}=\sum_{x \in\, \mathcal{C}}
S_{ax}S_{bx}S_{\overline{c}x}/S_{{\openone}x}$. 
Additionally, a theorem known as ``Ocneanu rigidity'' states that, given a set
of fusion rules, there are only finitely many corresponding TQFTs with these
rules \cite{Etingof05}. In other words, knowledge of the S-matrix is
sufficient to pin down the topological order of the state to a finite number
of possibilities. Clearly, it may be difficult to measure all elements of the
S-matrix by the methods described here. It appears particularly challenging to
invoke tunneling of anyonic charges different from that of the quasihole,
though one may speculate on
techniques that may eventually prove successful, such as resonant effects with
intermediate
antidots of tunable geometry and capacity on the tunneling arms.
Still, the S-matrix has many special properties and so even a partial
measurement of fairly low accuracy may be sufficient to determine it. In
addition to Eq.~(\ref{sprops}), any S-matrix must satisfy a set of constraints
coming from the Verlinde formula and
the fact that the fusion coefficients are integers. Also, the first row
of the S-matrix must be real and positive, because of its relation to the quantum
dimensions (and in fact, all elements are numbers with special algebraic
properties). Finally, given an S-matrix, there must be a
diagonal matrix $T$ which together with $S$ generates a representation
of the modular group $\text{SL}(2,\mathbb{Z})$, implying that $(ST)^3=S^2$.
For any fixed number of particle types, only finitely many different
S-matrices are known (and it is conjectured that only finitely many exist). Hence,
once some the S-matrix elements and the number of different charges are known from
measurements, one may look at the finite list of known S-matrices and
hope to identify one that matches. In conclusion, for any two-dimensional system,
interference experiments as described here can in principle determine
the fusion rules and even a finite set of TQFTs, one of which will fully describe the
topological order.

\begin{acknowledgments}
The authors would like to thank E.~Ardonne, P.~Fendley, A.~Kitaev, C.~Nayak,
J.~Preskill,
and Z.~Wang for many illuminating discussions. This work was supported in part
by the NSF under Grant No.~EIA-0086038. P.~B.\ and K.~S.\ would like to
acknowledge the hospitality of Microsoft Project Q and KITP. K.~S.\ is also
grateful for the
hospitality of the IQI.
\end{acknowledgments}

\bibliography{../bibs/corr}

\end{document}